\documentclass{nature_modified}
\usepackage{amsmath, amssymb} 
\usepackage{graphicx}

\title{Observation of Aubry transition in finite atom chains via friction}

\author{Alexei Bylinskii$^{1,2}$, Dorian Gangloff$^{1,2}$, Ian Counts$^1$ and Vladan Vuleti\'c$^{1,3}$}

\begin{document}

\maketitle

\begin{affiliations}
 \item Massachusetts Institute of Technology
 \item equal contributions
 \item vuletic@mit.edu
\end{affiliations}

\begin{abstract}
The highly nonlinear many-body physics of a chain of mutually interacting atoms in contact with a periodic substrate gives rise to complex static and dynamical phenomena, such as structural phase transitions and friction. In the limit of an infinite chain incommensurate with the substrate, Aubry predicted a structural transition with increasing substrate potential, from the chain's intrinsic arrangement free to slide on the substrate, to a pinned arrangement favoring the substrate pattern\cite{AubryTransition1983,Aubry1983b,Peyrard2000}. To date, the Aubry transition has not been observed. Here, using a chain of cold ions subject to a periodic optical potential\cite{Garcia-Mata2006,Benassi2011,Pruttivarasin2011b,Karpa2013a,Mandelli2015a} we qualitatively and quantitatively establish a close relation between Aubry's sliding-to-pinned transition and superlubricity breaking in stick-slip friction\cite{Weiss1996,Weiss1997b}. Using friction measurements with high spatial resolution and individual ion detection\cite{Bylinskii2015,Gangloff2015}, we experimentally observe the Aubry transition and the onset of its hallmark fractal atomic arrangement. Notably, the observed critical lattice depth for a finite chain agrees well with the Aubry prediction for an infinite chain. Our results elucidate the connection between competing ordering patterns and superlubricity in nanocontacts - the elementary building blocks of friction.
\end{abstract}

\begin{figure}[h]
   \includegraphics*[scale=0.45]{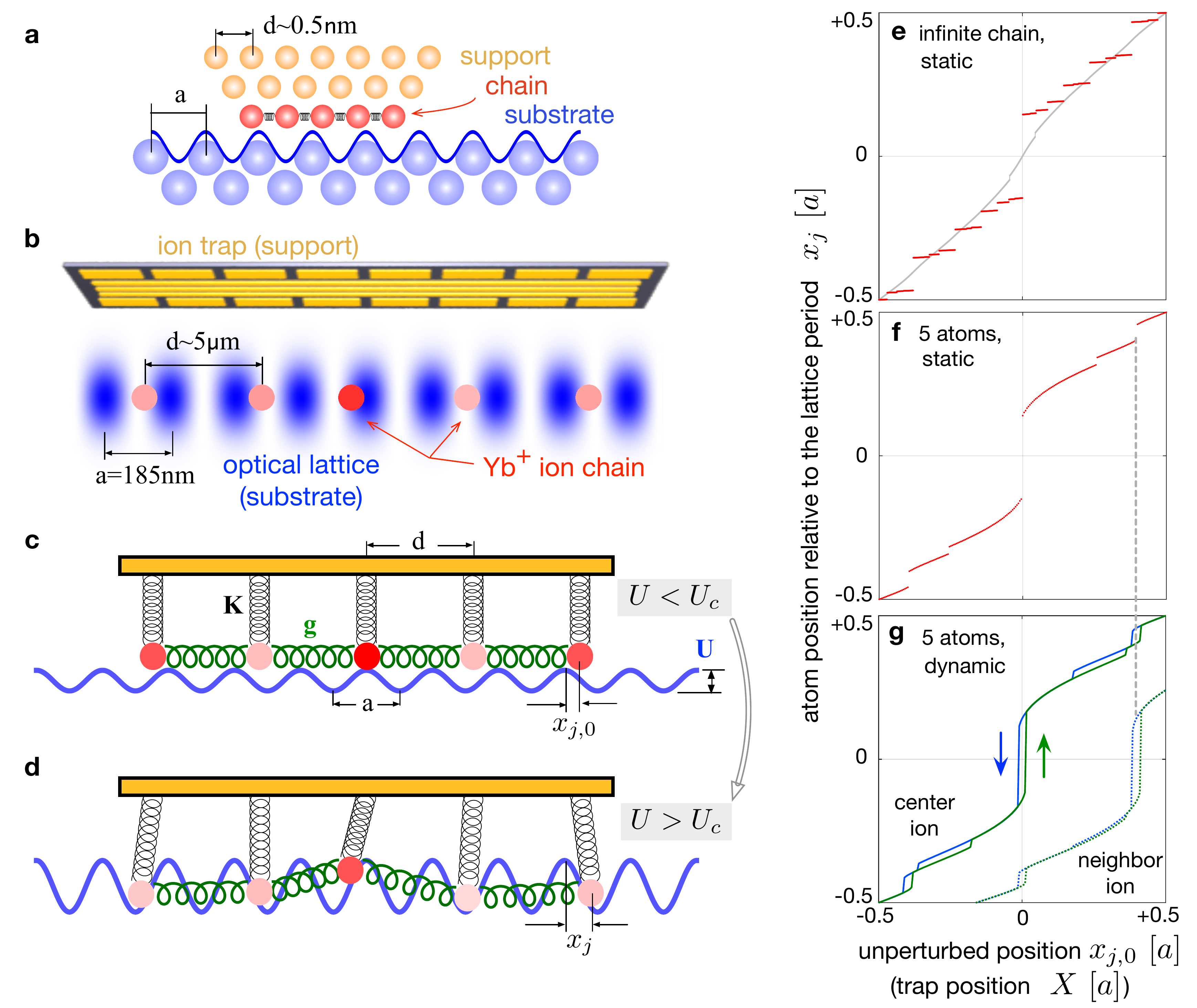}
   \caption{\textbf{A nanocontact, the ion-lattice system, the Frenkel-Kontorova-Tomlinson (FKT) model, and the Aubry transition.} \textbf{(a)} A few-atom nanocontact, such as occurs at a frictional interface, can be modeled as a contact-layer atomic chain (red), attached to a support (yellow), and interacting with the substrate lattice via a periodic potential (blue). \textbf{(b)} Our system\cite{Bylinskii2015,Karpa2013a,Cetina2013} of an ion chain trapped in a Paul trap (support) with typical axial vibrational frequency $\omega_0/(2\pi)\approx 360$ kHz and subject to an optical-lattice potential (substrate) with depth $U/k_B$ tunable between 50 $\mu$K and 2 mK. The ions are laser-cooled deep into the lattice to temperatures $\sim 0.05 U/k_B$. Each ion's position relative to the lattice is observed via the position-dependent ion fluorescence that results from the laser cooling scheme\cite{Bylinskii2015}. \textbf{(c,d)} The FKT model\cite{Weiss1996,Weiss1997b} of systems a) and b). Interatomic springs of stiffness $g$ couple neighboring atoms in a chain of period $d$, while external springs of stiffness $K$ couple them to the support. Each atom's position as measured from the nearest lattice maximum in the absence (presence) of the lattice is given by $x_{j,0}$ ($x_j$). Interaction with a sinusoidal potential of depth $U$ and period $a$ pins the chain above a critical lattice depth $U_c$, resulting in avoided position regions around lattice maxima. \textbf{(e)} In the limit $N\rightarrow\infty$, $K\rightarrow 0$ of the Frenkel-Kontorova model, and at $d($mod $a)/a=(\sqrt{5}-1)/2$ (the golden ratio), the curve of allowed ground state positions $x_{j}$ versus $x_{j,0}$ is continuous below the Aubry transition (grey) and forms a fractal staircase above the Aubry transition (red). (Numerical results shown are for  $N=101$ atoms and $g/K=128$.) \textbf{(f,g)} For a realistic nanocontact, for example $N=5$, $g/K=1$ and $d($mod $)a=(\sqrt{5}-1)/2$, a finite staircase forms above the transition. In a dynamical situation (fast translation of the support), the gaps in this staircase appear as hysteresis loops in measurements of friction above the superlubricity-breaking transition. The trajectory for a neighboring ion is shown with a dotted line to emphasize that the secondary gaps are due to the primary instabilities of other ions in the chain near their respective lattice maxima.
   }
       \label{fig:AubryFig1}
\end{figure}

The static arrangement of a chain of interacting atoms subject to the periodic potential of a substrate lattice is governed by the competition between the associated length and energy scales: the intrinsic chain spacing competes against the lattice spacing, and the elastic energy of the chain competes against the pinning energy of the lattice potential. Aubry found a remarkable transition\cite{AubryTransition1983,Aubry1983b,Peyrard2000} resulting from this competition when considering an infinite chain of atoms joined by springs and subject to an incommensurate sinusoidal lattice potential (the Frenkel-Kontorova model\cite{BraunKivsharFKbook}, see Fig.~\ref{fig:AubryFig1}). Below a critical depth of the lattice potential $U_c$, chain stiffness resists pinning and favors the chain's intrinsic arrangement, resulting in a translationally invariant sliding phase. Above the critical depth, the lattice potential overcomes the stiffness of the chain, which becomes unstable against pinning. The atoms reorganize towards lattice minima and avoid lattice maxima, which become Peierls-Nabarro (PN) energy barriers. This results in analyticity-breaking in the atomic positions, which form an everywhere-discontinuous fractal pattern relative to the lattice period (Fig.\ref{fig:AubryFig1}e).

The problem of an atomic chain subject to a periodic potential is central to understanding nanocontacts between solids\cite{Mo2009} (Fig.\ref{fig:AubryFig1}a), composite materials\cite{Ward2015a}, dislocations in crystals\cite{BraunKivsharFKbook}, adsorbed monolayers\cite{Braun2006a}, and biomolecular transport\cite{Bormuth2009}. Realistically, those situations involve a finite atom chain, which might additionally be attached to an external support. Furthermore, dynamical phenomena in those systems, such as friction, are of tremendous practical importance, yet remain poorly understood even at the few-atom level. In particular, stick-slip friction, which represents the dominant mode of energy dissipation and wear at the nanoscale, originates from chain pinning, and is thought to be intimately related to Aubry's pinned phase\cite{Shinjo1993}, where the force required to move the chain over the finite PN barriers corresponds to the friction force. Hence, below a critical lattice depth, such that PN barriers are absent, Aubry's sliding phase should manifest dynamically in smooth translation of the chain over the lattice, known as superlubricity\cite{Shinjo1993} - nearly frictionless transport as a result of vanishing stick-slip friction. Microscopic studies of friction\cite{Vanossi2013,Krylov2014a} have been performed experimentally with tools ranging from very sensitive nanotribology apparatuses\cite{Carpick1997,Szlufarska2008,Urbakh2010} to colloidal crystals in optical lattices\cite{Bohlein2012}. Although signatures of superlubricity have been observed at the nanoscale\cite{Socoliuc2004,Dienwiebel2004}, to our best knowledge, a direct and quantitative connection of these observations to Aubry's sliding phase has never been established.

Following a number of proposals\cite{Garcia-Mata2006,Benassi2011,Pruttivarasin2011b}, we recently studied friction between chains of trapped Yb$^+$ ions and an optical lattice (Fig.\ref{fig:AubryFig1}b) with atom-by-atom control unavailable in condensed-matter systems\cite{Bylinskii2015,Gangloff2015}. We showed that superlubricity can be achieved by structurally mismatching the ion chain to the optical lattice\cite{Bylinskii2015}, and, studying the velocity and temperature dependence of friction\cite{Gangloff2015}, identified a regime where stick-slip friction is minimally affected by finite temperature. In the present work, operating in the said regime, we observe a transition from superlubricity to stick-slip with increasing lattice depth. Simultaneously to this superlubricity breaking we observe a formation of discontinuities in the allowed ground-state ion positions, corresponding to Aubry's analyticity breaking. We show that the critical lattice depth of this transition in the few-ion chains agrees with Aubry's analytic result in the infinite-chain limit extended to include finite external confinement\cite{Weiss1996,Weiss1997b}. Thus, we observe the Aubry transition in a finite system, and establish qualitatively and quantitatively that the Aubry transition and the onset of stick-slip friction represent static and dynamic aspects, respectively, of the same physical phenomenon.

In our system\cite{Bylinskii2015,Karpa2013a,Cetina2013}, the atomic chain is a self-organized one-dimensional Coulomb crystal of laser-cooled $^{174}$Yb$^+$ ions spaced by a few micrometers in a linear Paul trap. The periodic lattice potential, with lattice constant $a=185$ nm, is due to an optical dipole force on the atoms by a standing wave of light (Fig.\ref{fig:AubryFig1}b). At finite depth $U$ the lattice deforms the ion arrangement, which is specified by the position $x_j$ of each ion $j$ measured from the nearest lattice maximum (Fig.\ref{fig:AubryFig1}d). $x_{j,0}$ corresponds to the intrinsic arrangement in the limit $U=0$ of an unperturbed chain (Fig.\ref{fig:AubryFig1}c). Although the intrinsic ion spacing $d$ is not uniform along the chain, it can be controlled with nanometer precision via the Paul trap's axial confinement to match or mismatch the chain to the lattice. The chain is matched when all neighboring ions are separated by an integer number of lattice periods $d ($mod $a) = 0$. The chain is maximally mismatched when the sum of lattice forces on the ions at their intrinsic positions $x_{j,0}$ cancels\cite{Bylinskii2015,Igarashi2008}: $\sum_j $sin$\big(2\pi x_{j,0}+\phi\big)=0$ for any $\phi$. This system is well described by a generalized Frenkel-Kontorova-Tomlinson (FKT) model\cite{Weiss1996,Weiss1997b} of a nanocontact, involving an $N$-atom chain joined by interatomic springs of stiffness (spring constant) $g$ (given here by Coulomb forces), and attached to a rigid support by external springs of stiffness $K=m\omega_0^2$, where $\omega_0/2\pi$ is the axial common-mode vibrational frequency in the Paul trap, and $m$ is an ion's mass (Fig.\ref{fig:AubryFig1}c). Both components of chain stiffness $g$ and $K$ cause resistance to deformation and pinning by the lattice when the ion chain is mismatched.

In a weak lattice, the ion chain is deformed but still continuously distributed relative to the lattice, i.e. the ions assume all positions relative to the lattice as the support (Paul trap) is translated along the lattice. In a sufficiently deep lattice, ions become pinned when they are excluded from regions near lattice maxima as a result of anti-confinement by the negative lattice potential curvature $\frac{2\pi^2}{a^2}U$ overcoming the confinement due to total chain stiffness ($g$ and $K$), and creating finite PN barriers there (Fig.\ref{fig:AubryFig1}d). This anti-confinement of a given ion near a lattice maximum, combined with interatomic forces further excludes other ions from other position ranges in the lattice. Thus, analyticity breaking is the formation of discontinuities in the curve of an atom's position $x_{j}$ (relative to the lattice) versus its unperturbed position $x_{j,0}$ as the support is translated. The primary discontinuity is a gap near $x_{j,0}=0$ from $x_j <0$ on one side of the PN barrier to $x_j>0$ on the other side, and progressively smaller gaps form near $x_{j,0}= d_{ij}($mod $a)$ due to anti-confinement of progressively further neighbors $i$ in the chain at intrinsic distances $d_{ij}$ away. For a finite mismatched chain of $N$ ions, each curve $x_j$ should accordingly have $N$ gaps as the support is translated by one lattice period $a$ (Fig.\ref{fig:AubryFig1}f). In the $N=\infty$ theoretical limit, for an incommensurate chain with irrational $d ($mod $a) /a$, the curve of $x_j$ (which coincides with the hull function\cite{BraunKivsharFKbook}) forms a nowhere-analytic fractal staircase with an infinite number of gaps (Fig.\ref{fig:AubryFig1}e). The primary gap parameterizes the Aubry analyticity-breaking transition\cite{Coppersmith1983}. In a finite chain, the primary gap for the center ion corresponds to displacement of this ion and the chain to one side of the emerging central PN barrier, parameterizing a reflection-symmetry-breaking transition that represents a finite version of the Aubry transition\cite{Benassi2011,Sharma1984,Braiman1990,BraunKivsharFKbook}.

The appearance of finite PN barriers, which give rise to the gaps in the static ground-state chain arrangement at the Aubry transition, leads in the dynamic situation to stick-slip friction associated with the now bistable PN potential. Consider the ions' relative positions $x_{j}$ versus the support position $X$ (or, equivalently, versus the intrinsic positions $x_{j,0}(t)=x_{j,0} +X(t)$) when the support is moved sufficiently quickly so that the chain does not have time to thermally relax across the PN barrier to the ground-state global minimum\cite{Gangloff2015}. Then, the chain sticks to the metastable minimum below the static position gap, until the PN barrier vanishes, and a slip to the global minimum above the position gap occurs (Fig.\ref{fig:AubryFig1}f). When the support is moved in the opposite direction, the chain sticks to the metastable minimum above the position gap before slipping back to the global minimum below the position gap. This dynamical process results in hysteresis loops enclosing the gaps in the static arrangement (Fig.\ref{fig:AubryFig1}g). Below the pinning transition, there are no PN barriers: the superlubric chain always follows the global minimum and the dynamic curves of $x_j$ coincide with the continuous static curves. Thus, the hysteresis that can be used to measure friction\cite{Bylinskii2015,Gangloff2015} across the superlubricity-breaking transition, can also be used to measure the opening of gaps in the atomic position distribution across the analyticity-breaking transition: the two are the dynamic and static aspects, respectively, of the Aubry transition.

\begin{figure}[h]
   \includegraphics*[scale=0.25]{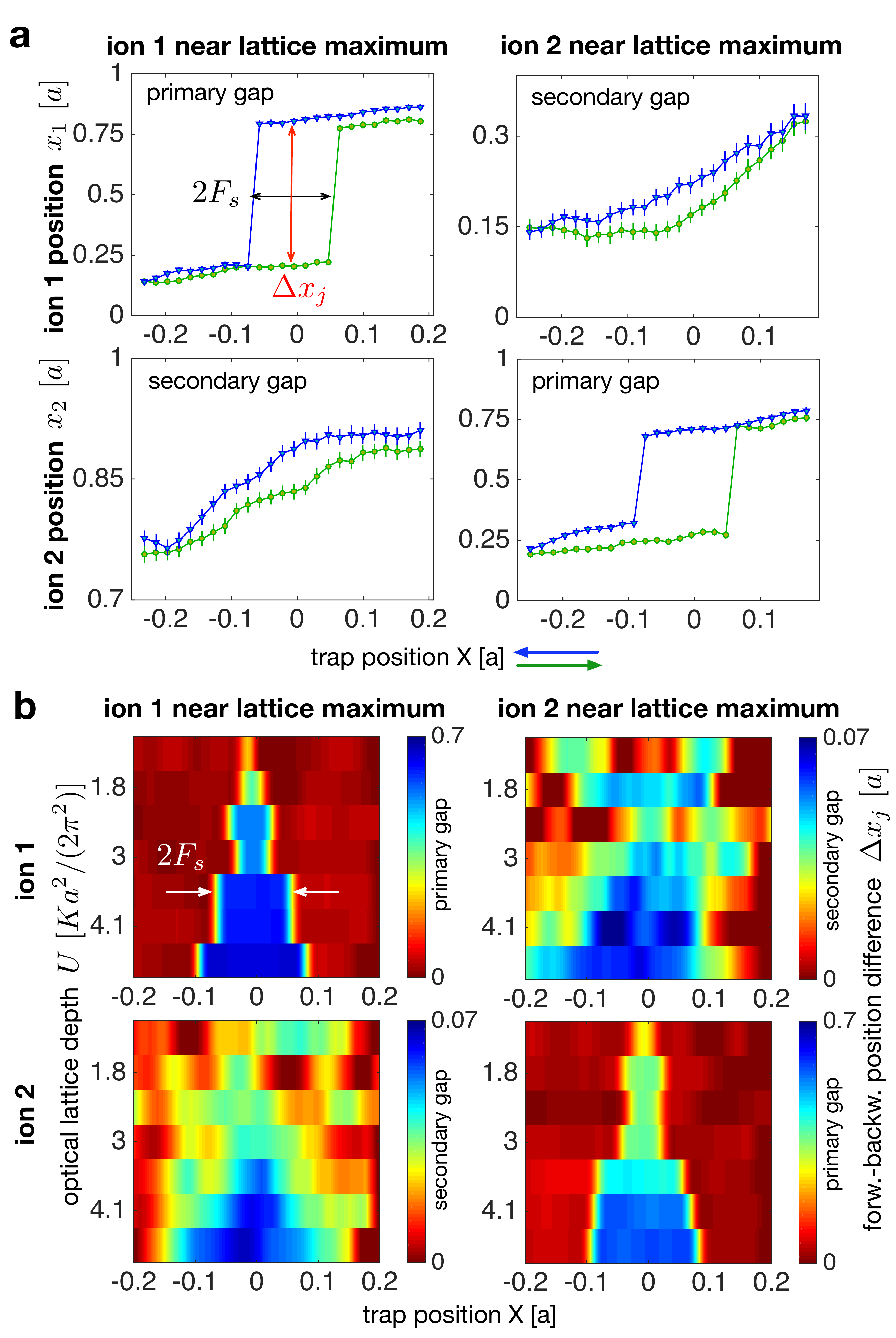}
   \caption{\textbf{Observation of primary and secondary position gap formation in a mismatched two-ion chain.} \textbf{(a)} The measured positions $x_1,x_2$ versus the unperturbed positions $x_{1,0},x_{2,0}$ show both a primary hysteresis loop\cite{Bylinskii2015} for each ion, and a secondary hysteresis loop induced by the other ion's hysteresis, when the trap position $X$ is quickly moved forward and backward in the vicinity of the corresponding Peierls-Nabarro barriers; here $x_{j,0}(t)=x_{j,0} +X(t)$.  These loops directly reveal the gaps in each ion's ground-state position distribution. The height of each loop measures the corresponding gap, and the width of the loops measures the static friction force $F_s$. The two ions in this measurement are mismatched to the lattice with $d($mod $a)/a\approx 2/3$. The secondary gap, corresponding to a much smaller energy scale, is smoothed due to finite ion temperature. The error bars show a statistical uncertainty of one standard deviation. \textbf{(b)} As the lattice depth $U$ is increased, the primary and secondary position gaps open up, as seen in the increasing difference between the forward and backward ion positions $\Delta x_j$, shown in grayscale, and stick-slip friction sets in, as manifested by the increasing static friction force, corresponding to the half-width of the darker region $\Delta x_j>0$. Note the different scales for the primary and secondary loops on the ion position axis in a) and on the color axis in b).}
       \label{fig:AubryFig2}
\end{figure}

We perform measurements on the system by applying an external electric force $F$ to quickly\cite{Bylinskii2015,Gangloff2015, supplementary} move back and forth the position $X(t)=F(t)/K$ of the axial trapping potential. The dynamic position curves $x_{j}$ are then reconstructed from the observed ion fluorescence\cite{Bylinskii2015}. In the elementary case of a mismatched two-ion chain we observe two hysteresis loops for each ion: a large one due to the primary slips of the ion over its lattice maximum, and a smaller one due to secondary slips induced by the primary slips of the other ion (Fig.\ref{fig:AubryFig2}a). These hysteresis loops correspond to the appearance of gaps in the allowed ion positions for the finite chain. The heights of the loops $\Delta x_j$ give the desired static gaps (discontinuities) in the atomic position distribution. The left and right edges of the loops correspond to the slipping events in the stick-slip process, and the separation between them equals twice the static friction force $F_s$ required to pull the chain over the corresponding PN barrier\cite{Bylinskii2015}. The area enclosed by the loops is the energy dissipated by stick-slip in the two slip events\cite{Bylinskii2015}. We observe these loops, and correspondingly the position gap $\Delta x_j$ and the static friction force $F_s$, to grow with increasing lattice depth (Fig.\ref{fig:AubryFig2}b) as a result of growing PN barriers.

\begin{figure}[h]
   \includegraphics*[scale=0.75]{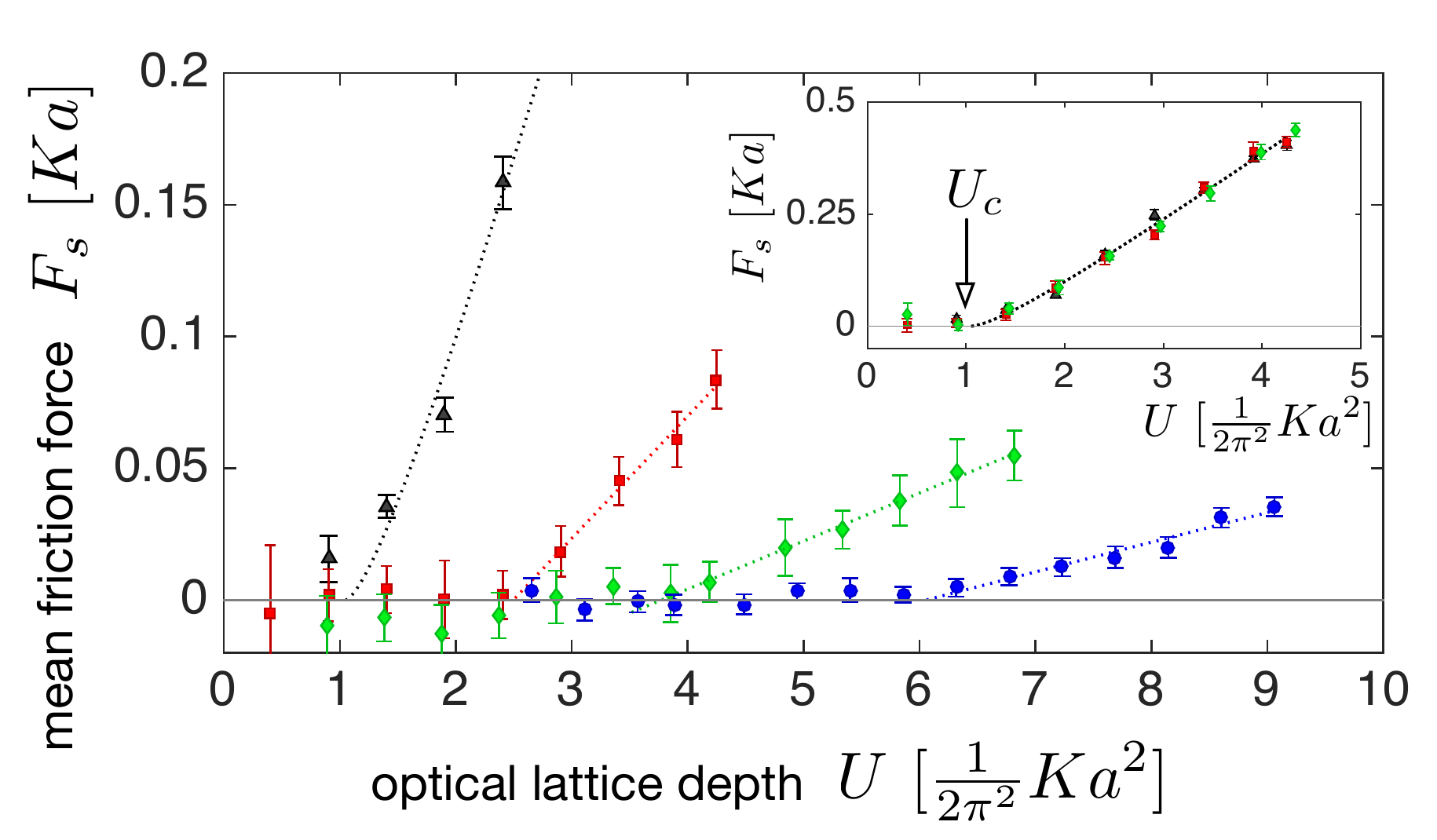}
       \centering
   \caption{\textbf{Observation of superlubricity breaking for matched and mismatched ion chains.} For $N=1-5$ ions matched to the lattice (inset), the measured mean friction force per ion as a function of lattice depth is independent of ion number, corresponding to the single-ion Prandtl-Tomlinson (PT) model (black dashed line), with superlubricity for $U<U_c=Ka^2/(2\pi^2)$; the interatomic springs $g$ do not have any effect. For $N=$ 1 (black), 2 (red), 3 (green) and 5 (blue) ions mismatched to the lattice (main figure), to break superlubricity the lattice must also overcome the interatomic spring stiffness $g$, which increases with N, thus extending the superlubric regime to larger values of $U_c$. The error bars represent statistical uncertainties of one standard deviation. To extract the critical value $U_c$ from the data, we fit an analytical formula for the PT model to the single-ion and the matched chain results (black dotted line). For the mismatched chain results, we fit a lowest-order piecewise linear model $c_0 + c_1(U-U_c)\cdot H(U-U_c)$ where $H$ is the Heaviside step function (red, green and blue dotted lines).
   }
       \label{fig:AubryFig3}
\end{figure}

For chains of $N$=1-5 ions, the measured static friction force $F_s$ reveals that the hysteresis loops open up at a finite lattice depth $U_c$ separating an analytic ($\Delta x_j=0$) and superlubric ($F_s=0$) phase for $U<U_c$ from a gapped phase ($\Delta x_j>0$) with finite friction ($F_s>0$) for $U>U_c$ (Fig.\ref{fig:AubryFig3}). When the ions are matched to the lattice, analyticity and superlubricity are observed to break at $U_c\approx Ka^2/(2\pi^2)$ for all $N$ (Fig.\ref{fig:AubryFig3} inset) as in the single-atom limit of the FKT model corresponding to the Prandtl-Tomlinson model\cite{Vanossi2013,Socoliuc2004}. In other words, for a matched chain anti-confinement by the lattice only needs to overcome the external chain stiffness $K$, as the interatomic springs remain at constant length, and play no role. When the ions are mismatched to the lattice, analyticity and superlubricity are observed to break at values of $U_c>Ka^2/(2\pi^2)$ (Fig.\ref{fig:AubryFig3}) as interatomic springs of stiffness $g$ help keep the ions near their unperturbed positions against the lattice forces in the static limit, and store some of the lattice potential energy during motion. $U_c$ is observed to increase with ion number as a result of increasing effective interatomic spring stiffness $g$. This stiffness is calculated by linearizing next-neighbor Coulomb forces for small lattice-induced deformations $\delta d/d<a/d\ll 1$, and these forces increase as ion separations $d$ decrease when more ions are loaded in the harmonic Paul trap, leading to $g/K\approx \frac{1}{4}N^{1.65}$ at the center of the chain\cite{supplementary}. Thus, in the large-$N$ limit, the interatomic springs dominate the chain stiffness and the pinning behavior, and the system is well described by the Frenkel-Kontorova model\cite{Garcia-Mata2006}. In this limit, and when the chain and the substrate are maximally incommensurate at a spacing ratio equal to the golden ratio $d($mod $a)/a=(\sqrt{5}-1)/2$, the Aubry transition should occur at the largest value $U_c=ga^2/(2\pi^2)$. Finite external confinement $K$ of an infinite golden-ratio ratio chain should increase this bound\cite{Weiss1996,Weiss1997b} to $U_c\gtrsim(g+K)a^2/(2\pi^2)$.

\begin{figure}[h]
   \includegraphics*[scale=0.6]{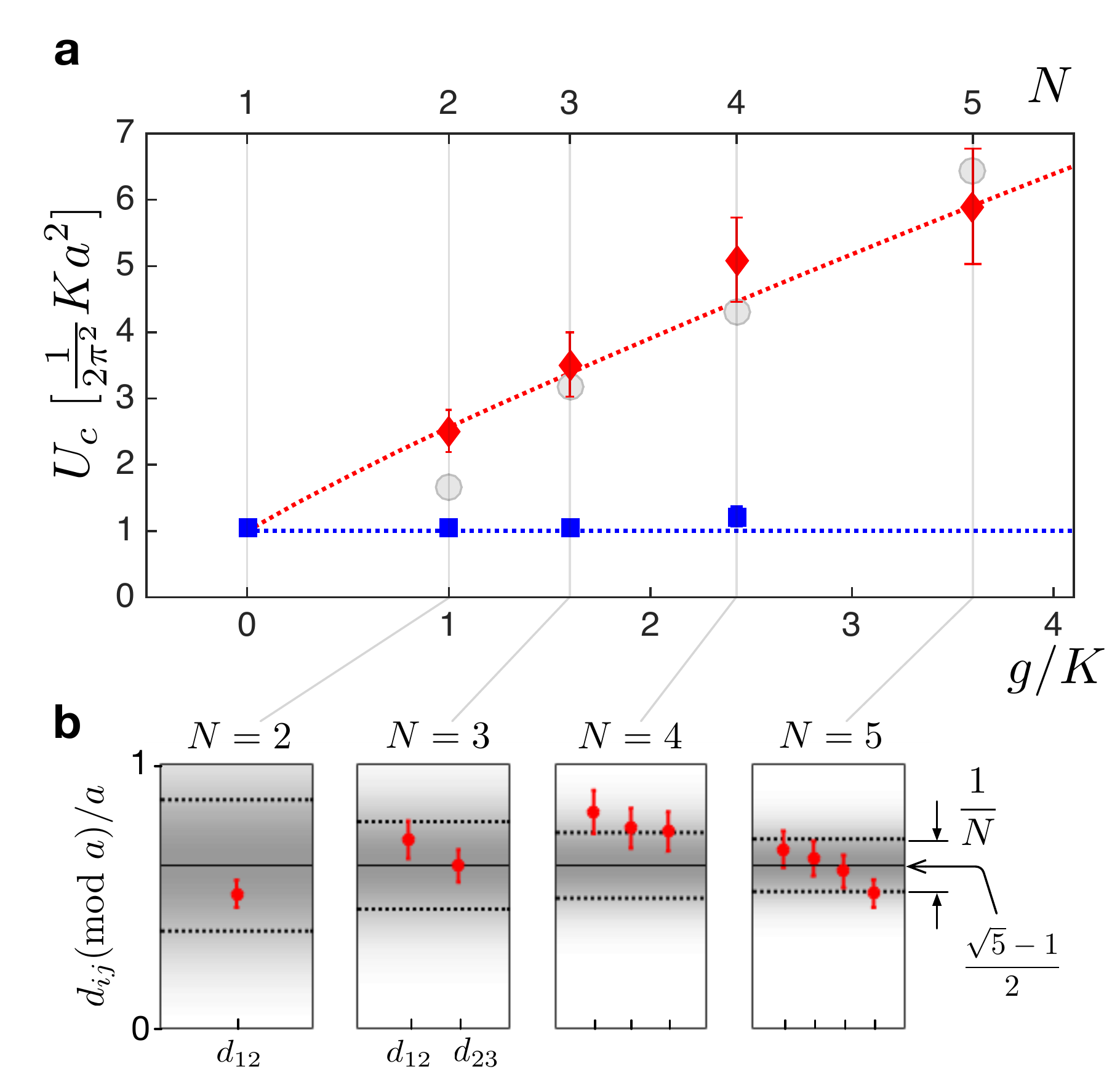}
       \centering
   \caption{\textbf{Aubry transition in finite mismatched ion chains.} \textbf{(a)} The measured critical value $U_c$ at the chain center is plotted against the interatomic spring stiffness $g$ normalized to the fixed external spring stiffness $K$ ($\sim 1.5\times 10^{-12}$ N/m). Each value of $g/K$ is obtained at a different ion number $N$ plotted on the second horizontal axis at the top. In the matched case (blue filled squares), regardless of ion number or interatomic springs stiffness, the critical lattice depth is given by the PT limit $U_c=Ka^2/(2\pi^2)$ (blue dotted line). In the mismatched case (red filled diamonds), the critical lattice depth follows the Aubry transition ($N=\infty$, $d($mod $a)/a=(\sqrt{5}-1)/2$), modified\cite{Weiss1996,Weiss1997b} by the finite external confinement $K$ (red dotted line). The gray filled circles are zero-temperature numerical simulations of our finite, inhomogeneous system with Coulomb interactions. The error bars represent a $\pm12\%$ systematic uncertainty in applying the lowest-order fitting model to extract $U_c$ from the data. \textbf{(b)} Measured neighbor distances along each of the mismatched chains used in a). The distances are inhomogeneous due to the harmonic confinement of the Coulomb chain\cite{supplementary} and due to a 1\%-scale asymmetry of the harmonic trap. Despite these effects, the neighbor distances relative to the lattice period $d_{j,j+1}($mod $a)/a$ still fall within the $1/N$ resolution window (gray shading and dotted lines at $1 \sigma$) around the golden ratio $(\sqrt{5}-1)/2$.
   }
       \label{fig:AubryFig4}
\end{figure}

At the center of our finite chains, the measured values of $U_c$ at different values of $g/K$, obtained for chains of different $N$, coincide with the calculated values for the Aubry transition in an infinite golden-ratio chain with the corresponding values of $g/K$ (Fig.\ref{fig:AubryFig4}a). In other words, our finite inhomogeneous system simply tuned to cancel the lattice forces on an unperturbed arrangement of the chain is as robust against the breaking of analyticity and superlubricity as Aubry's upper bound in an infinite, maximally incommensurate chain. This can be explained by the finite periodicity resolution $1/N$ set by the size of a finite system, which prevents distinguishing these two cases. Indeed, for the mismatched chain arrangements that are observed to follow the Aubry upper bound, the intrinsic separations between neighboring ions $d_{j,j+1}($mod $a)/a$ fall within $1/N$ of the golden ratio (Fig.\ref{fig:AubryFig4}b). 

In this work we observe, with atom-by-atom control, the transition from superlubricity to stick-slip and from continuous to gapped position distributions in finite chains of atoms as a function of increasing interaction strength with a periodic substrate. We establish the connection between this transition, relevant for the interaction of real surfaces governed by finite-size nanocontacts, and Aubry's theoretical concept of analyticity breaking in infinite chains. Furthermore, this transition could become an exotic quantum phase transition in a regime with quantum tunneling of atoms between the lattice sites, possibly accessible in our system. This could lead to a quantum-mechanical picture of friction and of adsorbed monolayers, possibly relevant at the nanoscale and at cold surfaces.


\begin{methods}

\subsection{Position detection via fluorescence}

Our dynamic position curves $x_{j}$ are reconstructed from the observed ion fluorescence, which varies proportionally to the optical potential energy $\tfrac{1}{2}U\big(1+ \cos(2\pi x_{j})\big)$ experienced by the ion. This is a result of our laser-cooling scheme, which uses the optical lattice to couple the vibrational levels $n$ and $n - 2$ of the ion's quantized motion in the optical lattice well\cite{Karpa2013a}. The spatial dependence of this Raman coupling is such that the off-resonant transition $n \rightarrow n$, which on resonance would be stronger by two orders of the Lamb-Dicke factor $\eta$ ($\eta \approx 10\%$ for our system), increases from lattice node to lattice anti-node proportionally to the optical potential. The stronger this coupling is, the larger the scattered fluorescence, resulting in the position-dependent fluorescence signal, which, when time-resolved, amounts to sub-wavelength imaging of the ion's average trajectory.

\end{methods}




\begin{thebibliography}{10}
\expandafter\ifx\csname url\endcsname\relax
  \def\url#1{\texttt{#1}}\fi
\expandafter\ifx\csname urlprefix\endcsname\relax\def\urlprefix{URL }\fi
\providecommand{\bibinfo}[2]{#2}
\providecommand{\eprint}[2][]{\url{#2}}

\bibitem{AubryTransition1983}
\bibinfo{author}{Aubry, S.}
\newblock \bibinfo{title}{{The twist map, the extended Frenkel-Kontorova model
  and the devil's staircase}}.
\newblock \emph{\bibinfo{journal}{Physica D: Nonlinear Phenomena}}
  \textbf{\bibinfo{volume}{7}}, \bibinfo{pages}{240--258}
  (\bibinfo{year}{1983}).

\bibitem{Aubry1983b}
\bibinfo{author}{Aubry, S.} \& \bibinfo{author}{{Le Daeron}, P.}
\newblock \bibinfo{title}{{The discrete Frenkel-Kontorova model and its
  extensions}}.
\newblock \emph{\bibinfo{journal}{Physica D: Nonlinear Phenomena}}
  \textbf{\bibinfo{volume}{8}}, \bibinfo{pages}{381--422}
  (\bibinfo{year}{1983}).

\bibitem{Peyrard2000}
\bibinfo{author}{Peyrard, M.} \& \bibinfo{author}{Aubry, S.}
\newblock \bibinfo{title}{{Critical behaviour at the transition by breaking of
  analyticity in the discrete Frenkel-Kontorova model}}.
\newblock \emph{\bibinfo{journal}{Journal of Physics C: Solid State Physics}}
  \textbf{\bibinfo{volume}{16}}, \bibinfo{pages}{1593--1608}
  (\bibinfo{year}{2000}).

\bibitem{Garcia-Mata2006}
\bibinfo{author}{Garc\'{\i}a-Mata, I.}, \bibinfo{author}{Zhirov, O.~V.} \&
  \bibinfo{author}{Shepelyansky, D.~L.}
\newblock \bibinfo{title}{{Frenkel-Kontorova model with cold trapped ions}}.
\newblock \emph{\bibinfo{journal}{The European Physical Journal D}}
  \textbf{\bibinfo{volume}{41}}, \bibinfo{pages}{325--330}
  (\bibinfo{year}{2006}).

\bibitem{Benassi2011}
\bibinfo{author}{Benassi, A.}, \bibinfo{author}{Vanossi, A.} \&
  \bibinfo{author}{Tosatti, E.}
\newblock \bibinfo{title}{{Nanofriction in cold ion traps.}}
\newblock \emph{\bibinfo{journal}{Nature communications}}
  \textbf{\bibinfo{volume}{2}}, \bibinfo{pages}{236} (\bibinfo{year}{2011}).

\bibitem{Pruttivarasin2011b}
\bibinfo{author}{Pruttivarasin, T.}, \bibinfo{author}{Ramm, M.},
  \bibinfo{author}{Talukdar, I.}, \bibinfo{author}{Kreuter, A.} \&
  \bibinfo{author}{H\"{a}ffner, H.}
\newblock \bibinfo{title}{{Trapped ions in optical lattices for probing
  oscillator chain models}}.
\newblock \emph{\bibinfo{journal}{New Journal of Physics}}
  \textbf{\bibinfo{volume}{13}}, \bibinfo{pages}{075012}
  (\bibinfo{year}{2011}).

\bibitem{Karpa2013a}
\bibinfo{author}{Karpa, L.}, \bibinfo{author}{Bylinskii, A.},
  \bibinfo{author}{Gangloff, D.}, \bibinfo{author}{Cetina, M.} \&
  \bibinfo{author}{Vuleti\'{c}, V.}
\newblock \bibinfo{title}{{Suppression of Ion Transport due to Long-Lived
  Subwavelength Localization by an Optical Lattice}}.
\newblock \emph{\bibinfo{journal}{Physical Review Letters}}
  \textbf{\bibinfo{volume}{111}}, \bibinfo{pages}{163002}
  (\bibinfo{year}{2013}).

\bibitem{Mandelli2015a}
\bibinfo{author}{Mandelli, D.} \& \bibinfo{author}{Tosatti, E.}
\newblock \bibinfo{title}{{Nanophysics: Microscopic friction emulators}}.
\newblock \emph{\bibinfo{journal}{Nature}} \textbf{\bibinfo{volume}{526}},
  \bibinfo{pages}{332--333} (\bibinfo{year}{2015}).

\bibitem{Weiss1996}
\bibinfo{author}{Weiss, M.} \& \bibinfo{author}{Elmer, F.-J.}
\newblock \bibinfo{title}{{Dry friction in the Frenkel-Kontorova-Tomlinson
  model: Static properties}}.
\newblock \emph{\bibinfo{journal}{Physical Review B}}
  \textbf{\bibinfo{volume}{53}}, \bibinfo{pages}{7539--7549}
  (\bibinfo{year}{1996}).

\bibitem{Weiss1997b}
\bibinfo{author}{Weiss, M.} \& \bibinfo{author}{Elmer, F.-J.}
\newblock \bibinfo{title}{{Dry Friction in the Frenkel-Kontorova-Tomlinson
  Model: Dynamical Properties}}.
\newblock \emph{\bibinfo{journal}{Zeitschrift f\"{u}r Physik B Condensed
  Matter}} \textbf{\bibinfo{volume}{69}}, \bibinfo{pages}{55--69}
  (\bibinfo{year}{1997}).
\newblock \eprint{9704110}.

\bibitem{Bylinskii2015}
\bibinfo{author}{Bylinskii, A.}, \bibinfo{author}{Gangloff, D.} \&
  \bibinfo{author}{Vuletic, V.}
\newblock \bibinfo{title}{{Tuning friction atom-by-atom in an ion-crystal
  simulator}}.
\newblock \emph{\bibinfo{journal}{Science}} \textbf{\bibinfo{volume}{348}},
  \bibinfo{pages}{1115--1119} (\bibinfo{year}{2015}).

\bibitem{Gangloff2015}
\bibinfo{author}{Gangloff, D.}, \bibinfo{author}{Bylinskii, A.},
  \bibinfo{author}{Counts, I.}, \bibinfo{author}{Jhe, W.} \&
  \bibinfo{author}{Vuletic, V.}
\newblock \bibinfo{title}{Velocity tuning of friction with two trapped atoms}.
\newblock \emph{\bibinfo{journal}{Nature Physics}} \textbf{\bibinfo{volume}{advance
  online publication}} (\bibinfo{year}{2015}), \bibinfo{DOI}{doi:10.1038/nphys3459}.

\bibitem{BraunKivsharFKbook}
\bibinfo{author}{Braun} \& \bibinfo{author}{Kivshar}.
\newblock \emph{\bibinfo{title}{{The Frenkel-Kontorova Model: Concepts,
  Methods, and Applications}}} (\bibinfo{year}{2004}).

\bibitem{Mo2009}
\bibinfo{author}{Mo, Y.}, \bibinfo{author}{Turner, K.~T.} \&
  \bibinfo{author}{Szlufarska, I.}
\newblock \bibinfo{title}{{Friction laws at the nanoscale.}}
\newblock \emph{\bibinfo{journal}{Nature}} \textbf{\bibinfo{volume}{457}},
  \bibinfo{pages}{1116--9} (\bibinfo{year}{2009}).

\bibitem{Ward2015a}
\bibinfo{author}{Ward, A.} \emph{et~al.}
\newblock \bibinfo{title}{{Solid friction between soft filaments}}.
\newblock \emph{\bibinfo{journal}{Nature Materials}}  (\bibinfo{year}{2015}).

\bibitem{Braun2006a}
\bibinfo{author}{Braun, O.} \& \bibinfo{author}{Naumovets, A.}
\newblock \bibinfo{title}{{Nanotribology: Microscopic mechanisms of friction}}.
\newblock \emph{\bibinfo{journal}{Surface Science Reports}}
  \textbf{\bibinfo{volume}{60}}, \bibinfo{pages}{79--158}
  (\bibinfo{year}{2006}).

\bibitem{Bormuth2009}
\bibinfo{author}{Bormuth, V.}, \bibinfo{author}{Varga, V.},
  \bibinfo{author}{Howard, J.} \& \bibinfo{author}{Sch\"{a}ffer, E.}
\newblock \bibinfo{title}{{Protein friction limits diffusive and directed
  movements of kinesin motors on microtubules.}}
\newblock \emph{\bibinfo{journal}{Science (New York, N.Y.)}}
  \textbf{\bibinfo{volume}{325}}, \bibinfo{pages}{870--3}
  (\bibinfo{year}{2009}).

\bibitem{Shinjo1993}
\bibinfo{author}{Shinjo, K.} \& \bibinfo{author}{Hirano, M.}
\newblock \bibinfo{title}{{Dynamics of friction: superlubric state}}.
\newblock \emph{\bibinfo{journal}{Surface Science}}
  \textbf{\bibinfo{volume}{283}}, \bibinfo{pages}{473--478}
  (\bibinfo{year}{1993}).

\bibitem{Vanossi2013}
\bibinfo{author}{Vanossi, A.}, \bibinfo{author}{Manini, N.},
  \bibinfo{author}{Urbakh, M.}, \bibinfo{author}{Zapperi, S.} \&
  \bibinfo{author}{Tosatti, E.}
\newblock \bibinfo{title}{{Colloquium: Modeling friction: From nanoscale to
  mesoscale}}.
\newblock \emph{\bibinfo{journal}{Reviews of Modern Physics}}
  \textbf{\bibinfo{volume}{85}}, \bibinfo{pages}{529--552}
  (\bibinfo{year}{2013}).

\bibitem{Krylov2014a}
\bibinfo{author}{Krylov, S.~Y.} \& \bibinfo{author}{Frenken, J. W.~M.}
\newblock \bibinfo{title}{{The physics of atomic-scale friction: Basic
  considerations and open questions}}.
\newblock \emph{\bibinfo{journal}{Physica Status Solidi (B)}}
  \textbf{\bibinfo{volume}{251}}, \bibinfo{pages}{711--736}
  (\bibinfo{year}{2014}).

\bibitem{Carpick1997}
\bibinfo{author}{Carpick, R.~W.} \& \bibinfo{author}{Salmeron, M.}
\newblock \bibinfo{title}{{Scratching the Surface: Fundamental Investigations
  of Tribology with Atomic Force Microscopy.}}
\newblock \emph{\bibinfo{journal}{Chemical reviews}}
  \textbf{\bibinfo{volume}{97}}, \bibinfo{pages}{1163--1194}
  (\bibinfo{year}{1997}).

\bibitem{Szlufarska2008}
\bibinfo{author}{Szlufarska, I.}, \bibinfo{author}{Chandross, M.} \&
  \bibinfo{author}{Carpick, R.~W.}
\newblock \bibinfo{title}{{Recent advances in single-asperity nanotribology}}.
\newblock \emph{\bibinfo{journal}{Journal of Physics D: Applied Physics}}
  \textbf{\bibinfo{volume}{41}}, \bibinfo{pages}{123001}
  (\bibinfo{year}{2008}).

\bibitem{Urbakh2010}
\bibinfo{author}{Urbakh, M.} \& \bibinfo{author}{Meyer, E.}
\newblock \bibinfo{title}{{Nanotribology: The renaissance of friction.}}
\newblock \emph{\bibinfo{journal}{Nature Materials}}
  \textbf{\bibinfo{volume}{9}}, \bibinfo{pages}{8--10} (\bibinfo{year}{2010}).

\bibitem{Bohlein2012}
\bibinfo{author}{Bohlein, T.}, \bibinfo{author}{Mikhael, J.} \&
  \bibinfo{author}{Bechinger, C.}
\newblock \bibinfo{title}{{Observation of kinks and antikinks in colloidal
  monolayers driven across ordered surfaces.}}
\newblock \emph{\bibinfo{journal}{Nature Materials}}
  \textbf{\bibinfo{volume}{11}}, \bibinfo{pages}{126--30}
  (\bibinfo{year}{2012}).

\bibitem{Socoliuc2004}
\bibinfo{author}{Socoliuc, A.}, \bibinfo{author}{Bennewitz, R.},
  \bibinfo{author}{Gnecco, E.} \& \bibinfo{author}{Meyer, E.}
\newblock \bibinfo{title}{{Transition from Stick-Slip to Continuous Sliding in
  Atomic Friction: Entering a New Regime of Ultralow Friction}}.
\newblock \emph{\bibinfo{journal}{Physical Review Letters}}
  \textbf{\bibinfo{volume}{92}}, \bibinfo{pages}{134301}
  (\bibinfo{year}{2004}).

\bibitem{Dienwiebel2004}
\bibinfo{author}{Dienwiebel, M.} \emph{et~al.}
\newblock \bibinfo{title}{{Superlubricity of Graphite}}.
\newblock \emph{\bibinfo{journal}{Physical Review Letters}}
  \textbf{\bibinfo{volume}{92}}, \bibinfo{pages}{126101}
  (\bibinfo{year}{2004}).

\bibitem{Cetina2013}
\bibinfo{author}{Cetina, M.} \emph{et~al.}
\newblock \bibinfo{title}{{One-dimensional array of ion chains coupled to an
  optical cavity}}.
\newblock \emph{\bibinfo{journal}{New Journal of Physics}}
  \textbf{\bibinfo{volume}{15}}, \bibinfo{pages}{053001}
  (\bibinfo{year}{2013}).

\bibitem{Igarashi2008}
\bibinfo{author}{Igarashi, M.}, \bibinfo{author}{Natori, A.} \&
  \bibinfo{author}{Nakamura, J.}
\newblock \bibinfo{title}{{Size effects in friction of multiatomic sliding
  contacts}}.
\newblock \emph{\bibinfo{journal}{Physical Review B - Condensed Matter and
  Materials Physics}} \textbf{\bibinfo{volume}{78}}, \bibinfo{pages}{1--5}
  (\bibinfo{year}{2008}).

\bibitem{Coppersmith1983}
\bibinfo{author}{Coppersmith, S.~N.} \& \bibinfo{author}{Fisher, D.}
\newblock \bibinfo{title}{{Pinning transition of the discrete sine-Gordon equation}}.
\newblock \emph{\bibinfo{journal}{Physical Review B}}
  \textbf{\bibinfo{volume}{28}}, \bibinfo{pages}{2566} (\bibinfo{year}{1983}).

\bibitem{Sharma1984}
\bibinfo{author}{Sharma, S.}, \bibinfo{author}{Bergersen, B.} \&
  \bibinfo{author}{B., J.}
\newblock \bibinfo{title}{{Aubry transition in a finite modulated chain}}.
\newblock \emph{\bibinfo{journal}{Physical Review B}}
  \bibinfo{pages}{6335--6340} (\bibinfo{year}{1984}).

\bibitem{Braiman1990}
\bibinfo{author}{Braiman, Y.}, \bibinfo{author}{Baumgarten, J.},
  \bibinfo{author}{Jortner, J.} \& \bibinfo{author}{Klafter, J.}
\newblock \bibinfo{title}{{Symmetry-Breaking Transition in Finite
  Frenkel-Kontorova Chains}}.
\newblock \emph{\bibinfo{journal}{Physical Review Letters}}
  \textbf{\bibinfo{volume}{65}}, \bibinfo{pages}{2398--2401}
  (\bibinfo{year}{1990}).

\bibitem{supplementary}
\bibinfo{title}{{Supplementary Materials}}.
\newblock \emph{\bibinfo{journal}{}}

\end{thebibliography}



\begin{addendum}
 \item[Acknowledgements] We thank Wonho Jhe for helpful discussions. A.B. and D.G. acknowledge scholarship support from NSERC. This work was supported by the NSF.
 \item[Author Contributions]
 A.B., D.G. and V.V. designed the experiments. D.G., A.B. and I.C. collected and analysed
data. All authors discussed the results and contributed to the manuscript preparation.
 \item[Competing Interests] The authors declare that they have no
competing financial interests.
 \item[Correspondence] Correspondence and requests for materials
should be addressed to Vladan Vuletic~(email: vuletic@mit.edu).
\end{addendum}


\onecolumn

\pagebreak
\twocolumn
{\LARGE Supplementary Materials}
\section{Effects of temperature on the Aubry transition.}

During our measurements of the ion position and of the static friction force, the ions have a finite temperature, estimated to be $0.05 U/k_B$ when the optical lattice depth $U$ is large ($\gg Ka^2/2\pi^2$). In this case, the effect of temperature can be nearly eliminated by moving the support faster than the thermal relaxation rate\cite{Bylinskii2015,Gangloff2015}. We show here that even when $U$ is low and near the Aubry transition point, where the PN energy barrier vanishes and the thermal relaxation rate is significant even at the fastest transport velocities, temperature only weakly affects the critical lattice depth $U_c$ extracted from our data. Instead, temperature mainly affects how steeply the friction force $F_s$ increases with lattice depth $U$ above $U_c$.

\begin{figure}[h]
   \includegraphics*[scale=0.7]{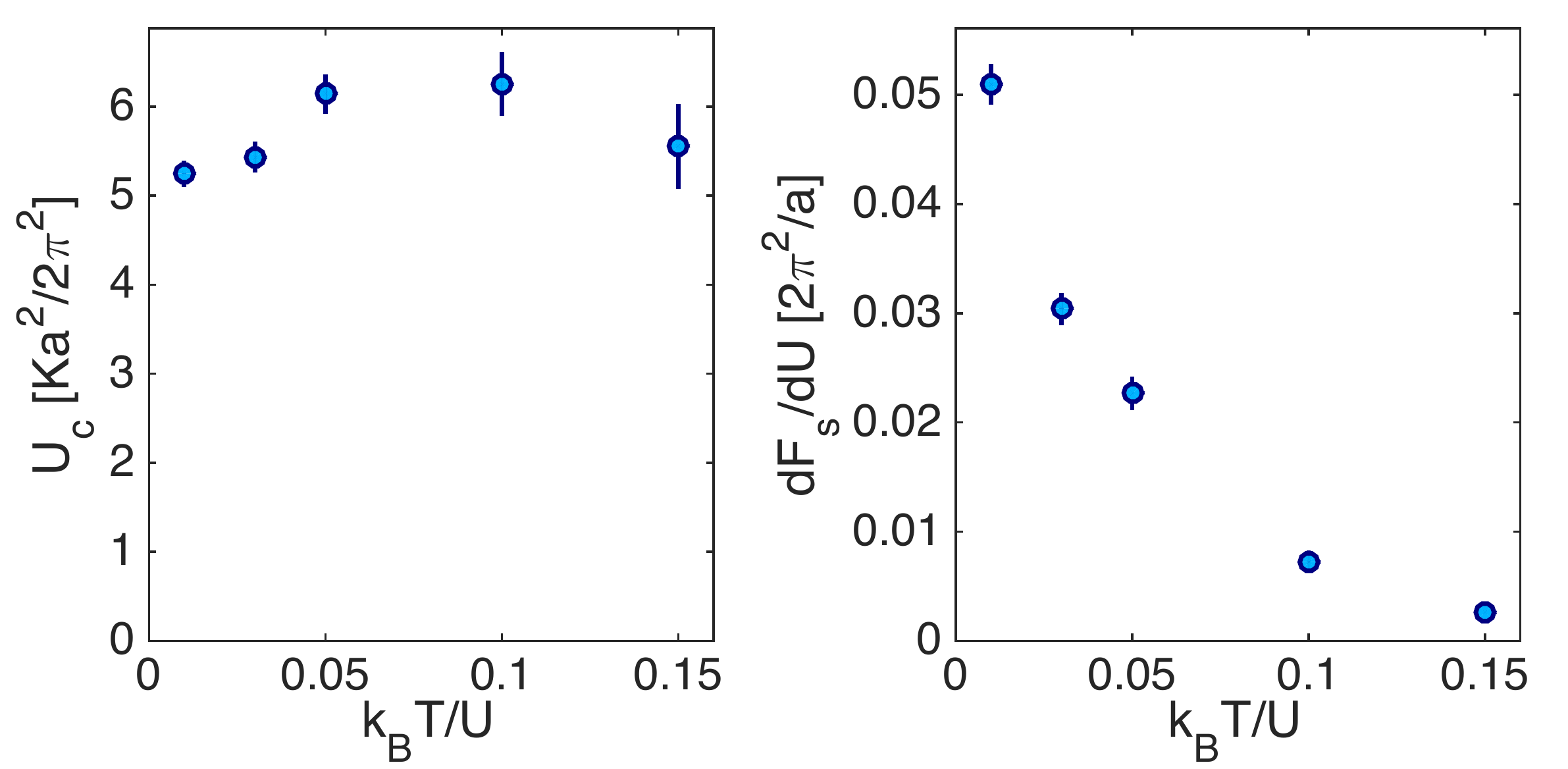}
   \caption{\textbf{Effects of temperature on the Aubry transition.} Dependence on temperature of the Aubry transition point $U_c$ (left) and of the friction onset $dF_s/dU$ (right) for a $N=5$ mismatched chain as obtained from numerical simulations of the full system dynamcis. Error bars represent a 68\% confidence interval on fitted values.
   }
       \label{fig:SuppFig1}
\end{figure}

To show this, we run a full dynamics simulation of our system, that is, an finite Coulomb chain with inhomogenous spacings. For the maximally mismatched chain of $N=5$ ions (the same arrangement as in the experiment), we extract the static friction force $F_s$ versus lattice depth $U$ for several temperatures in the range $0.01 - 0.15 U/k_B$, and extract the superlubricity-breaking point $U_c$ by fitting $F_s$ with a function of the form ${c_1\cdot(U-U_c)\cdot H(U-U_c)}$ as was done with the experimental data ($H$ is the Heaviside step function). The fitted values $U_c$ and $c_1=dF_s/dU$ are plotted against temperature in Fig.~\ref{fig:SuppFig1}.

We observe that for our experimentally measured temperature of $0.05 U/k_B$, the critical lattice depth $U_c$ is only pushed to higher values by $\sim$ 10\% relative to the $T=0$ limit. Interestingly, the slope $dF_s/dU$ is decreased by more than a factor of two for the same temperature difference, but this does not affect our conclusions about the observed Aubry transition.

\section{Coulomb forces as next-neighbor spring forces}

The harmonic confinement by external springs $K$ corresponds exactly to the axial confinement in the linear Paul trap at angular vibrational frequency $\omega_0 = \sqrt{K/m}$, which corresponds to the center-of-mass (COM) motion of the chain. We take the interatomic springs $g$ to correspond roughly to the highest-energy mode of the chain, for which the next-neighbors near the center of the chain move in anti-phase. This gives the interatomic spring stiffness as $g\approx \frac{1}{4}m\omega_{max}^2$. We show here why this is a valid approximation and how $g$ changes with $N$.

Consider a small displacement $\delta x$ of one of the ions in the chain on the order of the lattice constant $a$. This is small compared to the equilibrium spacing of the ions in the Paul trap, $\delta x \leq a \ll d_{ij}$ where $d_{ij}$ is the distance between any two ions $i$ and $j$ in the chain. One can then approximate the neighbor spring forces by linearizing the Coulomb forces around the equilibrium distance:


\begin{equation*}
\begin{split}
\delta F_{ij} &\approx  -2\frac{e^2}{4 \pi \epsilon_0 |d_{ij}|^3} \delta x
\end{split}
\end{equation*}

This gives the effective neighbor spring constant constant $g_{ij} = \frac{e^2}{2\pi \epsilon_0 |d_{ij}|^3}$. From this expression we see that next-neighbor interactions dominate the mean-field interaction of a single ion with the rest of the chain, even for an infinite homogeneous chain: the sum of forces from farther neighbors, as a fraction of the nearest neighbor forces, is $\sum_{n=2,\infty} \frac{1}{n^3} = \zeta(3)-1 \approx 20\%$ (where $\zeta$ is the Riemann zeta function).

In our inhomogeneous ion crystal, the largest next-neighbor spring constant $g_{j,j+1}$ is found for the center ion(s), where the next-neighbor separation is minimal. From numerical calculations, we find that this largest spring constant agrees well with $\frac{1}{4}m\omega_{max}^2$ even for a few-ion chain, and this is the value of $g$ we use when plotting $U_c$ against $g/K$ in Fig.~4a of the main text. For a fixed Paul trap confinement $\omega_0$, as ions are added to the trap, we find numerically that $\omega_{max}^2$ increases as $N^{1.64}$ due to closer confinement of neighboring ions in the center of the chain. As a result $g/K \approx \frac{1}{4}\left(\frac{\omega_{max}}{\omega_0}\right)^2 \approx \frac{1}{4} N^{1.64}$, allowing us to simply tune the FKT parameter $g/K$ with ion number $N$.

\section{Finite-system incommensurability}

In this section, we expand on the connection between incommensurate infinite chains and mismatched finite chains. An infinite homogeneous chain is maximally incommensurate with a substrate when the ratio of its next neighbor spacing $d_{j,j +1}$ to the substrate spacing $a$ is the golden ratio: ${(d_{j,j +1}\mod a) /a = (\sqrt{5}-1)/2}$. The mismatch of a finite chain to a periodic substrate is characterized by the matching parameter $q$, defined as the normalized barrier to translations of an infinitely rigid chain in that intrinsic arrangement\cite{Bylinskii2015,Igarashi2008}: $q=$max$_\phi\big[\frac{1}{N}\sum_j $sin$\big(2\pi x_{j,0}+\phi\big)\big]$. Thus, $q=1$ for a matched chain, which is equivalent to $d ($mod $a) = 0$, and $q=0$ for a maximally mismatched chain.

For a finite homogeneous chain of $N$ atoms, such that $N$ is a number in the Fibonacci sequence ($1,1,2,3,5,8,...$), one choice of chain spacing that maximally mismatches it with the substrate (i.e. $q=0$) is simply the $(i_N-1)$th rational approximation to the golden ratio (where $i_N$ is the index of $N$ in the Fibonacci series). For example, for $N=2$, $q=0$ for ${(d_{j,j +1}\mod a) /a = 1/2}$ which is the 2nd rational approximation to the golden ratio; for $N=3$, $q=0$ for ${(d_{j,j +1}\mod a) /a = 2/3}$ which is the 3rd rational approximation to the golden ratio; and for $N=5$, $q=0$ for ${(d_{j,j +1}\mod a) /a = 3/5}$ which is the 4th rational approximation to the golden ratio. These rational approximations are within $1/N$ of the true golden ratio $(\sqrt{5}-1)/2$. Within the discrete Fourier-limited resolution $1/N$ of a finite system of $N$ particles, this means that these spacings are practically indistinguishable from the golden ratio.

This result extends to arbitrary $N$. For a homogeneous chain of $N$ atoms, there exist $N - 1$ choices of spacing ${(d\mod a)/a = l/N}$ where ${1 \leq l < N}$ is an integer, such that the chain is mismatched ($q=0$) to the substrate\cite{Igarashi2008,Bylinskii2015}. It follows then that one can always find a value of $l$ such that $q=0$ and ${(d\mod a)/a = l/N}$ is within $1/(2N)$ of the golden ratio. Once again, this implies that this spacing is indistinguishable from the golden ratio within the Fourier-limited resolution $1/N$ in a finite chain. 

This reconciles the intuitions that a finite system is maximally superlubric when maximally mismatched to the substrate ($q=0$), and that an infinite system is maximally superlubric (statically, largest sliding phase) when maximally incommensurate with the substrate.

\onecolumn

\end{document}